# Gate-Tunable Resonant Tunneling in Double Bilayer Graphene Heterostructures

*Babak Fallahazad‡, Kayoung Lee‡, Sangwoo Kang‡, Jiamin Xue, Stefano Larentis, Christopher Corbet, Kyounghwan Kim, Hema C. P. Movva, Takashi Taniguchi [a), Kenji Watanabe [a),*

*Leonard F. Register, Sanjay K. Banerjee, Emanuel Tutuc\**

Microelectronics Research Center, Department of Electrical and Computer Engineering,

The University of Texas at Austin, Austin, TX 78758, USA

[a) National Institute for Materials Science, 1-1 Namiki Tsukuba Ibaraki 305-0044, Japan

**Keywords:** Bilayer graphene, hexagonal boron nitride, heterostructure, resonant tunneling, negative differential resistance, tunneling field-effect transistor.

**Abstract:** We demonstrate gate-tunable resonant tunneling and negative differential resistance in the interlayer current-voltage characteristics of rotationally aligned double bilayer graphene heterostructures separated by hexagonal boron-nitride (hBN) dielectric. An analysis of the heterostructure band alignment using individual layer densities, along with experimentally determined layer chemical potentials indicates that the resonance occurs when the energy bands of the two bilayer graphene are aligned. We discuss the tunneling resistance dependence on the interlayer hBN thickness, as well as the resonance width dependence on mobility and rotational alignment.

Recent progress in realization of atomically thin heterostructures by stacking two-dimensional (2D) atomic crystals, such as graphene, hexagonal boron nitride (hBN), and transition metal dichalcogenides (TMDs) has provided a versatile platform to probe new physical phenomena, and explore novel device functionalities [1], [2]. Combining such materials in vertical heterostructures may provide new insight into the electron physics in these materials through Coulomb drag [3], [4] or tunneling [5], [6]. Tunneling between two distinct 2D carrier systems, namely 2D-2D tunneling has been used in GaAs 2D electron [7], [8] and 2D hole systems [9]-[11] as a technique to probe the Fermi surface and quasi-particle lifetime. When the energy bands of two parallel 2D carrier systems are energetically aligned, momentum conserving tunneling leads to a resonantly enhanced tunneling conductivity and negative differential resistance (NDR).

The emergence of single or few atom-thick semiconductors, such as graphene and TMDs can open new routes to probe 2D-2D tunneling in their heterostructures, which in turn may enable new device applications [5], [12], [13]. While fascinating, resonant tunneling between two graphene or TMD layers realized using a layer-by-layer transfer approach is experimentally challenging because the energy band minima are located at the K points in the first Brillouin zone, and the large K-point momenta coupled with small rotational misalignment between the layers can readily obscure resonant tunneling.

Bilayer graphene consists of two monolayer graphene in Bernal stacking, and has a hyperbolic energy-momentum dispersion with a tunable bandgap [14]–[16]. Hexagonal boron-nitride is an insulator with an energy gap of 5.8 eV [17] and dielectric strength of 0.8 V/nm [18], which has emerged as the dielectric of choice for graphene [1] thanks to its atomically flat, and chemically inert surface. We demonstrate here resonant tunneling and NDR between two bilayer graphene flakes separated by an hBN dielectric. A detailed analysis of the band alignment in the heterostructure indicates that the NDR occurs when the charge neutrality points of the two layers are energetically aligned, suggesting momentum conserving tunneling is the mechanism responsible for the resonant tunneling.

Figure 1a shows a schematic representation of our double bilayer heterostructure devices, consisting of two bilayer graphene flakes separated by a thin hBN layer. The devices are fabricated through a sequence of bilayer graphene and hBN mechanical exfoliation, alignment, dry transfers, e-beam lithography, and plasma etching steps similar to the techniques reported in Refs. [1], [19]–[21] (supporting information). The bilayer graphene flakes selected for the device fabrication have at least one straight edge which is used as a reference to align the crystalline orientation of the bottom and top bilayer graphene during the transfer (Fig. 1b). The accuracy of the rotational alignment is mainly limited by the size of the flakes, and the resolution of the optical microscope. For a typical length of the bilayer graphene straight edge of $10 - 20$ μm, we estimate the rotational misalignment between the two bilayers in our devices to be less than 3 degrees. The interlayer hBN straight edges are not intentionally aligned with either the top or bottom graphene layers during transfers.

The interface between various materials in an atomically thin heterostructure plays a key role in device quality and tunneling uniformity. Particularly, the presence of contaminants, such as tape or resist residues, and wrinkles in the tunneling region and in between the layers changes the interlayer spacing and the local carrier density, which in turn makes the tunneling current distribution non-uniform. To achieve an atomically flat interface with minimum contamination, the heterostructure is annealed either after each transfer or after the stack completion in high vacuum ($10^{-6}$ Torr), at a temperature $T = 340°C$ for 8 hours. Figure 1c depicts the optical micrograph of the final device where the bottom and top bilayer graphene boundaries are marked by red and yellow dashed lines; the interlayer hBN is not visible in this micrograph. The devices are characterized at temperatures ranging from $T = 1.4$ K to room temperature, using small signal, low frequency lock-in techniques to probe the individual layer resistivities, and a parameter analyzer for the interlayer current-voltage characteristics. Eight devices were fabricated and investigated in this study; we focus here on data from three devices, labelled #1, #2, and #3. Device #1 consists of the double bilayer heterostructure separated by hBN where the top layer is exposed to ambient, while Devices #2 and #3 have the top layer capped with an additional hBN layer. The interlayer dielectric thickness of Devices #1, #2, and #3 correspond to six, five, and four hBN monolayers, respectively.

To characterize the double bilayer system, it is instructive to start with the characteristics of the individual layers. The device layout allows us to independently probe the bottom and top layer resistivites ($\rho_B$, $\rho_T$), and carrier densities ($n_B$, $n_T$) in the overlap (tunneling) region as a function of the back-gate ($V_{BG}$) and interlayer bias ($V_{TL}$) applied on the top layer; the bottom layer potential is kept at ground during all measurements. Figure 2 shows the bottom (panel a) and top (panel b) layer resistivity measured as a function of $V_{BG}$ and $V_{TL}$ in Device #1, at $T = 1.4$ K. The carrier mobility of Device #1 measured from the four-point conductivity is 150,000 - 160,000 cm$^2$/V·s for the bottom bilayer and 3,500 cm$^2$/V·s for the top bilayer at $T = 1.4$ K. The data of Fig. 2 indicate that the combination of gate biases at which both bilayer graphene are charge neutral, namely the double charge neutrality point (DNP), is: $V_{BG-DNP} = 20.2$ V and $V_{TL-DNP} = -0.235$ V.

At a given set of $V_{BG}$ and $V_{TL}$, the values of $n_B$ and $n_T$ can be calculated using the following equations [22]:

$$e(V_{BG} - V_{BG-DNP}) = \frac{e^2(n_B + n_T)}{C_{BG}} + \mu_B \qquad (1)$$

$$e(V_{TL} - V_{TL-DNP}) = -\frac{e^2 n_T}{C_{int}} + \mu_B - \mu_T \qquad (2)$$

Here $e$ is the electron charge, $C_{BG}$ is the back-gate capacitance, $C_{int}$ is the interlayer dielectric capacitance, $\mu_T$ and $\mu_B$ are the top and bottom bilayer graphene chemical potential measured with respect to the charge neutrality point, respectively. Solving eqs. 1 and 2 yields a one-to-one correspondence between the applied biases and the layer densities. Finding a self-consistent solution for eqs. 1 and 2 requires the $C_{BG}$ and $C_{int}$ values, and the layer chemical potential dependence on carrier density. We discuss in the following an experimental method to determine the capacitance values in a double bilayer graphene, along with the chemical potential dependence on the carrier density.

Along the charge neutrality line (CNL) of the top bilayer graphene [i.e. $n_T = \mu_T = 0$], eq. 2 reduces to $\mu_B = e(V_{TL} - V_{TL-DNP})$, thus the $\mu_B$ value at a given $V_{BG}$ can be determined along the top layer CNL. To determine the value of the $C_{BG}$, we measure $\rho_B$ and $\rho_T$ of the device in a perpendicular magnetic field. Figure 3a presents the $\rho_T$ contour plot of Device #1 measured as a function of $V_{BG}$ and $V_{TL}$, in a perpendicular magnetic field $B = 13$ T, and at $T = 1.5$ K. The charge neutrality line of the top bilayer graphene (dashed line in Fig. 3a) shows a staircase behavior, which stems from the bottom bilayer graphene chemical potential crossing the Landau levels (LLs) [19]. At a given LL filling factor ($v$), marked in Fig. 3a, the bottom bilayer graphene carrier density is $n_B = veB/h$; $h$ is the Planck constant. Writing eqs. 1 and 2 along the top bilayer CNL, combined with $n_B = veB/h$ yields:

$$C_{BG} = \frac{e^2 B}{h} \left( \frac{\Delta(V_{BG} - V_{TL})}{\Delta v} \right)^{-1} \qquad (3)$$

Where $\Delta(V_{BG} - V_{TL})$ is the change in $V_{BG} - V_{TL}$ corresponding to a bottom bilayer filling factor change $\Delta v$ along the top layer CNL (dashed line in Fig. 3a). Figure 3b shows a clear linear dependence of $(V_{BG} - V_{TL})$ vs. $v$, marked by circles in Fig. 3a. The slope of $(V_{BG} - V_{TL})$ vs. $v$ data along with eq. 3 yields $C_{BG}$ = 10.5 nF/cm$^2$ for Device #1, corresponding to 285 nm-thick SiO$_2$ in series with 40 nm-thick hBN dielectric.

The $n_B$ value along the top bilayer CNL can be calculated using eqs. 1 and 2:

$$n_B = \frac{C_{BG}}{e} \cdot [(V_{BG} - V_{BG-DNP}) - (V_{TL} - V_{TL-DNP})] \qquad (4)$$

Combining the $\mu_B$ values determined along the top layer CNL of Fig. 2b, with eq. 4 yields $\mu_B$ vs. $n_B$. Figure 3c shows $\mu_B$ vs. $n_B$ for Devices #1 and #3. We note that in addition to the layer densities, the applied $V_{BG}$ and $V_{TL}$ also change the transverse electric fields across the two layers (see supporting information). The chemical potential of the two devices match well at high carrier densities, but differ near $n_B = 0$ thanks to different transverse electric fields values across the bottom layer near the DNP [14]. Because the experimental data show the bilayer graphene chemical potential is weakly dependent on the transverse electric fields away from the neutrality point, and to simplify the solution of eqs. 1 and 2 we neglect the $\mu_T$ and $\mu_B$ dependence on the transverse electric field across the individual layers. The dashed line in the Fig. 3c depicts a polynomial fit to the experimental $\mu_B$ vs. $n_B$ data, which will be subsequently used to solve eqs. 1 and 2.

We now turn to the extraction of the $C_{int}$ value. Let us consider the bottom bilayer graphene CNL, marked by a dashed line in Fig. 2a. In a dual gated graphene device with metallic gates, the value of the top-gate capacitance can be readily extracted from the linear shift of the bottom graphene charge neutrality point with back-gate and top-gate voltages [23], which yields the top-gate to back-gate capacitance ratio. Because the top layer is not a perfect metal, using the slope of the bottom bilayer CNL of Fig. 2a to calculate $C_{int}$ neglects the contribution of the top bilayer quantum capacitance. Combining eqs. 1 and 2 along the bottom bilayer CNL, i.e. $n_B = \mu_B = 0$, we obtain the following expression that includes the quantum capacitance:

$$C_{int} = -\frac{eC_{BG} \cdot (V_{BG} - V_{BG-DNP})}{e(V_{TL} - V_{TL-DNP}) + \mu_T(C_{BG} \cdot (V_{BG} - V_{BG-DNP})/e)} \qquad (5)$$

Using eq. 5 and the $\mu_T$ vs. $n$ dependence of Fig. 3c, we determine an interlayer dielectric capacitance of $C_{int} = 1.02$ μF/cm$^2$ for Device #1. The $C_{int}$ values for Devices #2, and #3 are 1.23 μF/cm$^2$, and 1.55 μF/cm$^2$, respectively.

Now we turn to the interlayer current ($I_{int}$) - voltage characteristics of our devices. Figure 4a shows the $I_{int}$ vs. $V_{TL}$ for Device #1 measured at various $V_{BG}$ values, and at $T = 10$ K. For small bias values, $I_{int}$ increases monotonically with $V_{TL}$, corresponding to an interlayer resistance of 39 GΩ·μm$^2$. For $V_{BG}$ values ranging from 10 V to 30 V, the interlayer current-voltage traces show a marked resonance and NDR, which depend on the applied $V_{BG}$. Figure 4b presents the $I_{int}$ vs. $V_{TL}$ of Device #2 measured at room temperature. The normalized interlayer resistance of Device #2 at the limit of $V_{TL} = 0$ V is 1 GΩ·μm$^2$. Similar to the Device #1 data, we observe resonant tunneling and NDR in the interlayer current-voltage characteristics. A distinct difference between the two devices is that the resonance is centered around $V_{TL} = 0$ V in Device #2 by comparison to Device #1. As we discuss below, the NDR position can be explained quantitatively by considering the electrostatic potential across the double bilayer heterostructures.

Figure 4c shows the normalized interlayer resistance ($R_c$) in double bilayer graphene devices as a function of interlayer hBN thickness, from 4 to 8 monolayers, measured at zero interlayer bias, and at either low $T = 1.4$ - 20 K temperatures, or at room temperature. Data are included from both devices with and without resonant tunneling. The data show an exponential dependence on thickness of the tunneling barrier, similar to experimental tunneling data through hBN using graphite and gold electrodes [24]. These data indicate that the $R_c$ value is largely determined by the interlayer hBN thickness.

To better understand the origin of the observed NDR in Figs. 4a and 4b, it is instructive to examine the energy band alignment in the double bilayer graphene heterostructure. To determine if the NDR occurrence stems from momentum conserving tunneling, we examine the biasing conditions at which the charge neutrality points of the two bilayer graphene are aligned and the electrostatic potential drop across the interlayer dielectric is zero:

$$eV_{TL} + \mu_T(n_T) - \mu_B(n_B) = 0 \qquad (6)$$

Figure 5a illustrates the energy band alignment of a double bilayer graphene device at biasing conditions where the charge neutrality points of top and bottom bilayers are aligned, the condition most favorable for momentum conserving tunneling. The schematic ignores the band-gap induced in the two layers as a result of finite transverse electric fields (see supporting information), as the layer chemical potentials are controlled mainly by the carrier densities (Fig. 3c). The symbols in Fig. 5b show the experimental values of the tunneling resonance as a function of $V_{TL}$ and $V_{BG}$ for Devices #1 and #2, defined as the maximum conductivity point in Fig. 4a,b data. The solid lines show the calculated $V_{TL}$ vs. $V_{BG}$ values corresponding to layer densities and chemical potential that satisfy eq. 6, corresponding to the charge neutrality points of the two layers being aligned. The good agreement between the experimental values and calculations in Fig. 5b strongly suggests that the tunneling resonance occurs when the charge neutrality points of the two bilayer graphene are aligned, which in turn maximizes momentum (**k**) conserving tunneling between the two layers [25]–[27]. This observation is also in agreement with the findings in other 2D-2D systems where resonant tunneling occurs when the energy bands of the two quantum wells are aligned [8]–[11]. We note however, that in addition to the tunneling resonances, both devices exhibit a non-resonant tunneling current background which increases with $V_{TL}$, associated with non-momentum-conserving tunneling. This non-resonant tunneling component has a weak temperature dependence, which implies it is not caused by phonon assisted tunneling or thermionic emission.

Figure 5c shows the layer densities $n_T$ vs. $n_B$ calculated in Devices #1 and #2 at the tunneling resonance position corresponding to Figs. 4a,b data. In Device #1 the top (bottom) bilayer is populated with holes (electrons) at the tunneling resonance. In Device #2 the top bilayer is close to neutrality, while the bottom bilayer carrier type can be either hole or electron depending on the applied $V_{BG}$. Most notably, in both devices the tunneling resonance occurs at a fixed top layer density value. This observation can be understood using eq. 2 and eq. 6, which yield a fixed top layer density $n_T = (V_{TL-DNP} \cdot C_{int})/e$ when the charge neutrality points are aligned, independent of $V_{BG}$.

In addition to the location of the resonances, we also considered their broadening. Potential sources of broadening include finite initial and final state lifetimes $\tau$ due to scattering, rotational misalignment $\theta$, or the non-uniformity of tunneling associated with spatial inhomogeneities. While a detailed theoretical description of the tunneling in double bilayers is outside the scope of this study, in the following we provide estimates for the broadening associated with these mechanisms gauges in terms of the alignment of the band structures, i.e., the electrostatic potential difference between bilayers $V_{ES} = V_{TL} + [\mu_T(n_T) - \mu_B(n_B)]/e$.

The contribution from the carrier scattering lifetime ($\tau$) in either layer to the broadening width in units of volts is $\Delta V_\tau \cong \hbar/(e\tau)$, where $\hbar$ is the reduced Planck constant. Using the momentum relaxation time $\tau_m$ obtained from the carrier mobility $\mu = e\tau_m/m^*$, where $m^*$ is the effective mass, a lower limit of the broadening can be estimated to be $\Delta V_\tau \cong \hbar/(\mu m^*)$ . The broadening width associated with rotational misalignment can be estimated using the wave-vector difference $\Delta k = |\mathbf{K}|\theta$ illustrated in Fig. 6b, which translates into a broadening $\Delta V_\theta \cong \hbar \bar{v}|\mathbf{K}|\theta/e$, where $\bar{v}$ is an average velocity of the tunneling carriers, and $|\mathbf{K}| = 1.7 \times 10^{10}$ m$^{-1}$ is the wave-vector magnitude at the valley minima. Using the Fermi velocity of monolayer graphene $v_F$ = $1.1{\times}10^8$ cm/s as reference leads to a numerical expression $\Delta V_\theta \cong (215\, mV)(\bar{v}/v_F)(\theta/1º)$. The lower carrier velocity in bilayer by comparison to monolayer graphene leads to a reduced resonance broadening at a given rotational misalignment angle $\theta$. Moreover, a smoother resonance broadening shape for rotational misalignment is expected for the double bilayer graphene thanks to the quasi-parabolic energy-momentum dispersion, compared to a rotationally misaligned double graphene monolayer [5], [26].

The effective mass of bilayer graphene is both density and transverse electric field dependent [14], [19]. Using an average effective mass value $m^* = 0.05m_e$ [19], where $m_e$ is the bare electron mass, the lower layer mobility value in Device #1 of 3,500 cm$^2$/V·s corresponds to a broadening $\Delta V_\tau = 7$ mV. For Device #2 the corresponding broadening is $\Delta V_\tau = 11$ mV, using the top and bottom layer mobility values of 14,800 and 2,400 cm$^2$/V·s, respectively, measured at room temperature.

The experimental values for the tunneling resonance width are $\Delta V_{ES} \cong 12$ mV and $\Delta V_{ES} \cong 76$ mV for Devices #1 and #2, measured at $T = 10$ K, and room temperature respectively. These values are determined by fitting Lorentzian peaks to the $I_{int}$ data of Fig. 4a,b data plotted as a function of $V_{ES}$; an example is shown in Fig. 6c. Fitting a Lorentzian peak to the $I_{int}/V_{TL}$ vs. $V_{ES}$ data yields very similar $\Delta V_{ES}$ values. As the $\Delta V_\tau$ values calculated above are lower than the experimental values $\Delta V_{ES}$, we conclude that the broadening is mainly limited by rotational alignment in our devices, with Device #1 having a better alignment than Device #2. Although we cannot quantify experimentally the rotational misalignment in the two devices, we note that during fabrication Device #1 was annealed after each graphene and hBN layer transfer, while Device #2 was annealed after the double bilayer stack was completed. We speculate that multiple annealing steps may improve the rotational alignment between the layers.

In summary, we present a study of interlayer electron transport in double bilayer graphene. In devices where the bilayers straight edges were rotationally aligned during the fabrication we observe marked resonances in interlayer tunneling. Using individual layer densities and experimental values of the layer chemical potential we show that the resonances occur when the charge neutrality points of the two layers are energetically aligned, consistent with momentum-conserving tunneling. The interlayer conductivity values show an exponential dependence of the interlayer hBN thickness, and can serve to benchmark switching speed for potential device applications.

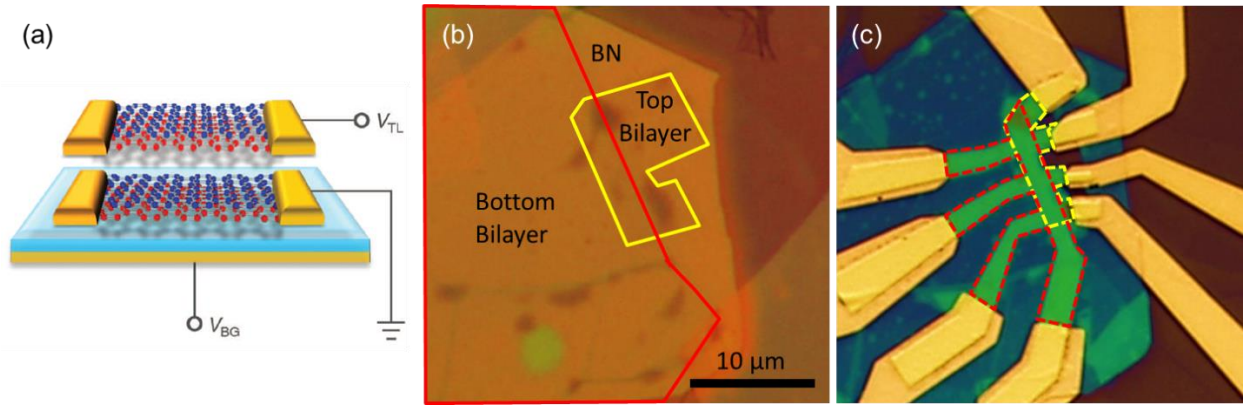

**Figure 1. Double bilayer device structure.** (a) Schematic of the double bilayer graphene device. (b) Optical micrograph of the top and bottom graphene flakes illustrating the alignment of straight edges. The red (yellow) lines mark the boundaries of the bottom (top) bilayer graphene. (c) Optical micrograph of the device. The red (yellow) dashed lines mark the bottom (top) bilayer graphene.

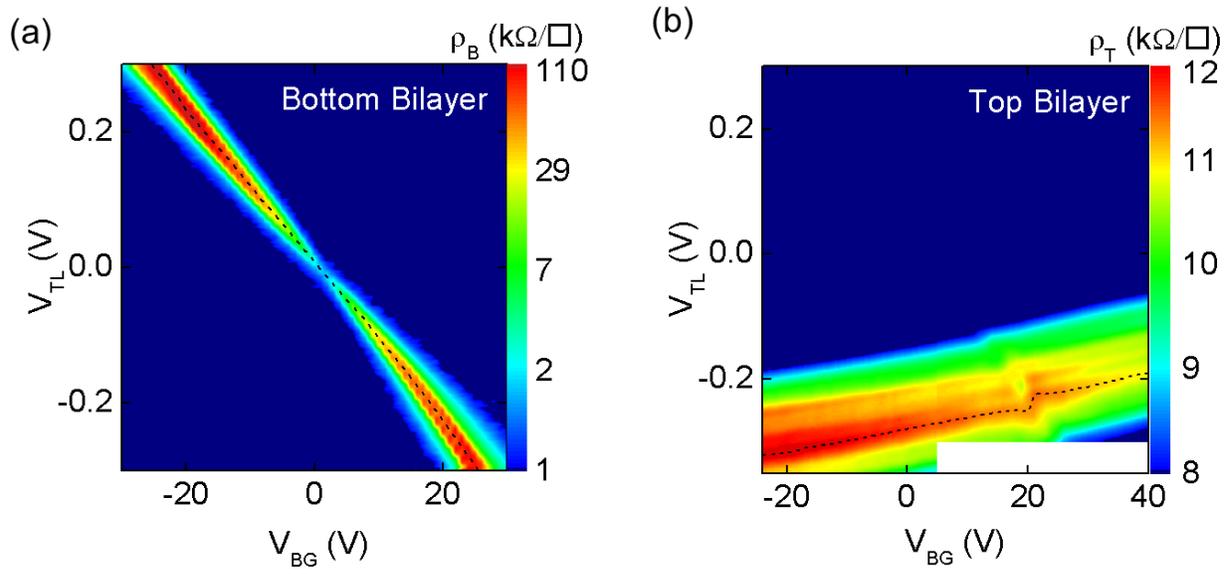

**Figure 2. Individual layer characterization.** Device #1 bottom [panel (a)] and top [panel (b)] bilayer graphene resistivity contour plots measured as a function of $V_{BG}$ and $V_{TL}$ at $T = 1.4$ K. The charge neutrality points in both panels are marked by black dashed lines.

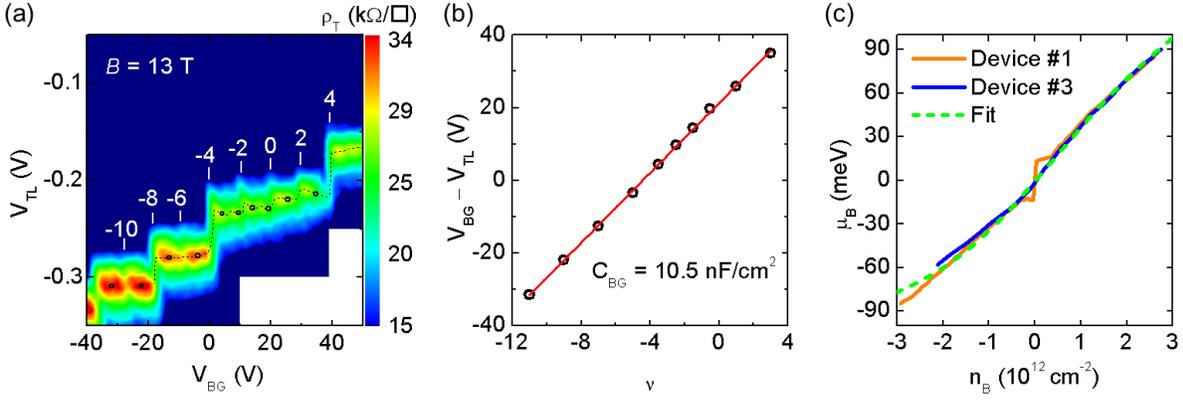

**Figure 3. Capacitance and chemical potential measurement.** (a) Contour plot of $\rho_T$ measured as a function of $V_{BG}$ and $V_{TL}$, at $B$ = 13 T and $T$ = 1.5 K in Device #1. The bottom bilayer graphene LL filling factors are marked. (b) $V_{BG} - V_{TL}$ vs. $\nu$ of the bottom bilayer showing a linear dependence; the $C_{BG}$ value is determined from the slope. (c) $\mu_B$ vs. $n_B$ for Devices #1 and #3. The dashed line is the polynomial fit to the experimental data.

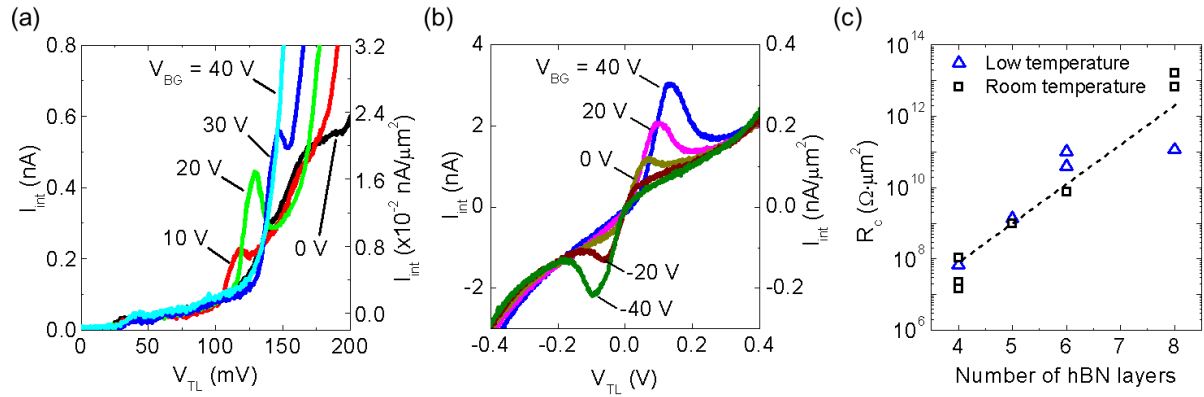

**Figure 4. Interlayer current-voltage characteristics and resonant tunneling.** $I_{int}$ vs. $V_{TL}$ of (a) Device #1 measured at $T = 10$ K, and (b) Device #2 measured at room temperature. The right axes in panels (a) and (b) show the interlayer current normalized by the active area. (c) Normalized interlayer resistance vs. number of hBN layers measured in multiple devices and at a low temperature of $T = 1.4 - 20$ K and at room temperature. The dashed line is a guide to the eye.

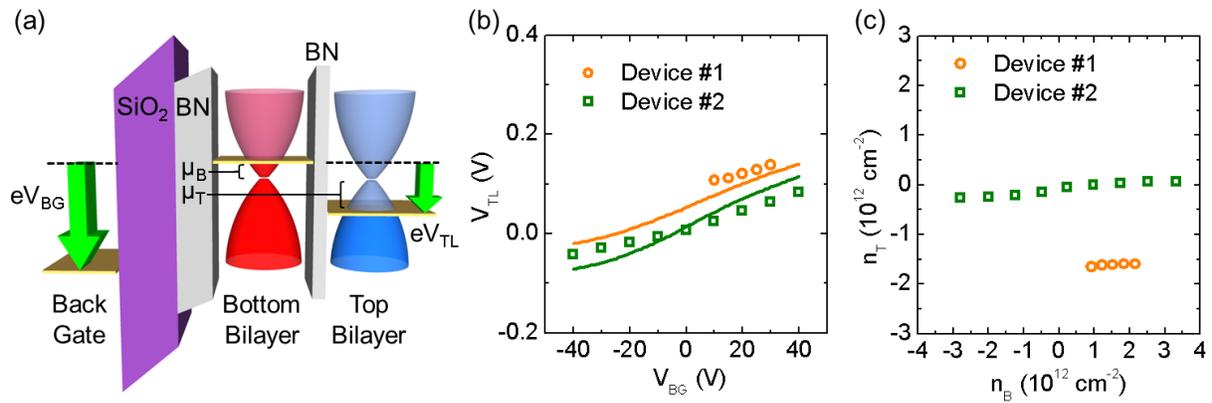

**Figure 5. Energy band alignment and carrier densities at tunneling resonance.** (a) Energy band diagram of the double bilayer graphene device when charge neutrality points of top and bottom bilayers are aligned. (b) $V_{TL}$ vs. $V_{BG}$ of Devices #1 and #2 at tunneling resonance (circles) and when charge neutrality points are aligned (solid line) (c) $n_T$ vs. $n_B$ of Devices #1 and #2 at tunneling resonance.

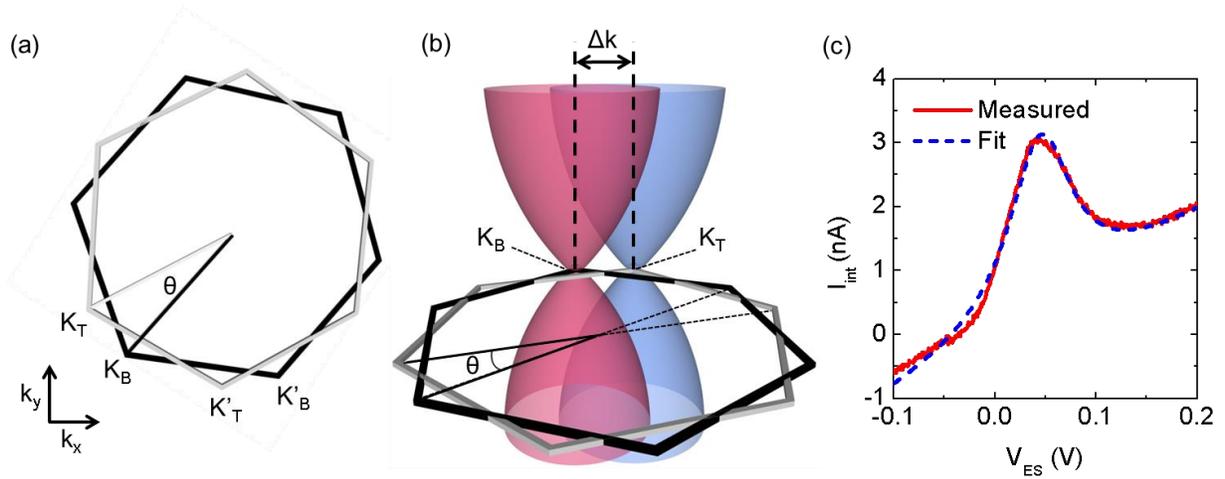

**Figure 6. Energy band diagram of rotationally misaligned bilayers.** (a) Brillouin zone boundaries of two hexagonal lattices rotationally misaligned by $\theta°$ in real space. $K_B$ ($K_T$) is the valley minimum the bottom (top) bilayer graphene. (b) A rotational misalignment by a small angle $\theta$ translates into valley separation in momentum space by $\Delta k \cong |\mathbf{K}|\theta$. (c) $I_{int}$ vs. $V_{ES}$ for Device #2 at $V_{BG} = 40$ V (solid line), along with a Lorentzian fit to the experimental data (dashed line) superimposed to a linear background.

ASSOCIATED CONTENT

**Supporting Information**

Details of device fabrication methods and *E*-field calculation.

AUTHOR INFORMATION

**Corresponding Author**

*[etutuc@mer.utexas.edu](mailto:etutuc@mer.utexas.edu);

**Author Contributions**

The manuscript was written through contributions of all authors. All authors have given approval to the final version of the manuscript. ‡These authors contributed equally.

**Funding Sources**

This work has been supported by NRI-SWAN, ONR, and Intel.

ACKNOWLEDGMENT

We thank Randall M. Feenstra for technical discussions.

# Supporting Information

**Device fabrication**

The fabrication starts with exfoliation of hBN on a silicon wafer covered with 285 nm-thick thermally grown $SiO_2$. Topography and thickness of the exfoliated hBN flakes are measured with atomic force microscopy (AFM), and flakes with minimum surface roughness and surface contamination are selected. On a separate silicon wafer covered with water soluble Polyvinyl Alchohol (PVA) and Poly(Methyl Methacrylate) (PMMA), bilayer graphene is mechanically exfoliated from natural graphite and identified using optical contrast and Raman spectroscopy. The PVA is dissolved in water, and the PMMA/bilayer graphene stack is transferred onto hBN flake using a thin glass slide. The PMMA film is then dissolved in acetone and the bilayer graphene is trimmed using EBL and $O_2$ plasma etching. Similarly, a thin hBN ($t_{hBN}$ = 1.2-1.8 nm) flake exfoliated on a PMMA/PVA/Si substrate is transferred onto the existing bilayer graphene. A second bilayer graphene is transferred onto the stack, and trimmed on top of the bottom bilayer graphene using EBL and $O_2$ plasma etching. Finally, metal contacts to both top and bottom bilayer graphene are defined through EBL, electron-beam evaporation of Ni and Au, and lift-off.

Device #2 is fabricated using the dry transfer method described in ref. [S1]. The device fabrication starts with mechanical exfoliation of bilayer graphene and hBN on $SiO_2$/Si substrate. Then, we spin coat poly-propylene carbonate (PPC) on a 1 mm-thick Polydimethylsiloxane (PDMS) film bonded to a thin glass slide. The glass/PDMS/PPC stack is used to pick up the top bilayer graphene, the thin interlayer hBN ($t_{hBN}$ = 1.2 nm), and the bottom bilayer graphene consecutively from SiO$_2$/Si substrates using the Van der Waals force between the two-dimensional crystals. The entire stack is transferred onto an hBN flake previously exfoliated on SiO$_2$/Si substrate. Figure 1(b) shows the transferred stack on top of bottom hBN/SiO$_2$/Si substrate. After dissolving the PPC, a sequence of EBL, O$_2$ and CHF$_3$ plasma etching is used to define the active area. Finally, the metal contacts are defined by EBL, e-beam evaporation of Ti-Au, and lift-off.

### Transverse electric field across the individual bilayers

The momentum-conserving tunneling between two bilayer graphene depends on their energy-momentum dispersion, and density of states. The band structure of bilayer graphene, particularly close to the CNP, can be tuned by an applied transverse electric ($E$) field, as a result of the applied $V_{BG}$ and $V_{TL}$. It is therefore instructive to examine the $E$-field value for the two bilayers in a double bilayer graphene heterostructure. The general expressions for transverse $E$-field across the top ($E_T$) and bottom ($E_B$) bilayers in a double bilayer graphene device are:

$$E_B = \frac{en_B}{2\varepsilon_0} + \frac{en_T}{\varepsilon_0} + E_{B0} \quad (S1)$$

$$E_T = \frac{en_T}{2\varepsilon_0} + E_{T0} \quad\quad (S2)$$

Here $n_T$ and $n_B$ are the top and bottom layer densities, respectively, and $\varepsilon_0$ is the vacuum permittivity. $E_{T0}$ and $E_{B0}$ are the transverse $E$-fields across the top and bottom bilayer at the DNP, as a result of unintentional layer doping. At a given $V_{BG}$ and $V_{TL}$, the $n_B$ and $n_T$ values can be calculated from eqs. 1 and 2. The $E_{B0}$ value can be calculated as following. We first determine $E_B = 0$ point, marked by minimum $\rho_B$ along the CNL of the bottom bilayer resistivity contour plot (Fig. S1a). At $E_B = 0$, eq. 1 and S1 yield:

$$E_{B0} = \frac{C_{BG}\Delta V_{BG}}{\varepsilon_0} \qquad (S3)$$

Here $\Delta V_{BG} = V_{BG-DNP} - V_{BG-E_B=0}$.

Finding the value of the $E_{T0}$ in a back-gated double bilayer device requires an assumption about the dopant position that cause the device DNP to shift from $V_{BG} = V_{TL} = 0$ V. To calculate the $E_{T0}$ in our devices assume the dopants are placed on the top bilayer graphene, an assumption most plausible when the top bilayer is uncapped, as in Device #1. Equation 1 combined with the Gauss law yield:

$$E_{T0} = \frac{C_{BG}V_{BG-DNP}}{\varepsilon_0}$$

Figures S1b and S1c show the calculated $E_T$ and $E_B$ in Device #1 and #2 along the locus of aligned neutrality points in the two bilayers, i.e. at the tunneling resonance, as a function of $V_{BG}$. At the tunneling resonance $E_B$ shows a linear dependence on $V_{BG}$, while $E_T$ remains constant. For Device #1, the condition $E_T = E_B$, desirable for identical energy-momentum dispersion in the two bilayers occurs at $V_{BG} = 24\,V$, and a finite $E$-field. For Device #2, $E_T = E_B$ closer to zero, and at $V_{BG} = -7\,V$. Figures 4a and S1b data combined suggest the tunneling resonance in Device #1 is strongest in the vicinity of the $E_T = E_B$ point, where the band structures are closely similar for both top and bottom bilayers. The tunneling resonance in Device #2 occurs over a wider range of $V_{BG}$ where the difference between the $E_T$ and $E_B$ can be as large as 0.34 V/nm.

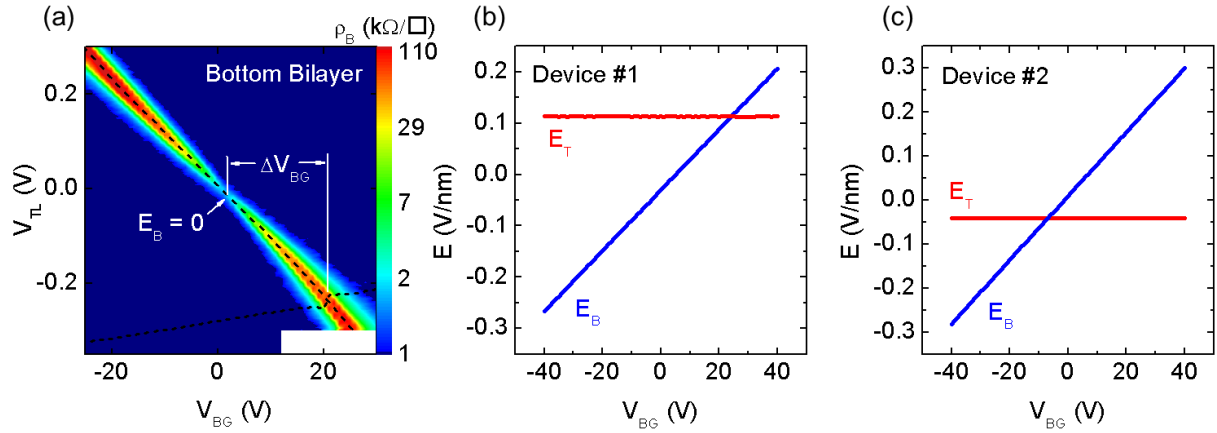

**Figure S1. Transverse *E*-fields across the top and bottom bilayers.** (a) Device #1 $\rho_B$ contour plot vs. $V_{BG}$ and $V_{TL}$, measured at $T = 1.4$ K. The CNL of the top bilayer graphene is added to mark the DNP. $E_T$ and $E_B$ in (b) Device #1, and (c) Device #2, calculated at the tunneling resonance.